\documentclass[nohyper,letterpaper,11pt,notoc]{JHEP3}
\usepackage{graphicx}
\usepackage{amssymb}
\usepackage{amsmath}
\usepackage{amsfonts}
\usepackage{latexsym}
\usepackage[all]{xy}
\usepackage{slashed}
\preprint{}

\newcommand{\gsim}{ \mathop{}_{\textstyle \sim}^{\textstyle >} }
\newcommand{\lsim}{ \mathop{}_{\textstyle \sim}^{\textstyle <}}

\newcommand{\tr}{\mathrm{tr}}
\newcommand{\mN}{m_{\scriptscriptstyle N}}
\newcommand{\ER}{E_{\scriptscriptstyle R}}
\newcommand{\med}{b}
\newcommand{\ald}{\alpha_{\scriptscriptstyle d}}
\newcommand{\gd}{g_{\scriptscriptstyle d}}

\newcommand{\be}{\begin{eqnarray}}
\newcommand{\ee}{\end{eqnarray}}

\newcommand{\cm}{\rm \, cm}

\title{PAMELA, DAMA, INTEGRAL and Signatures of Metastable Excited WIMPs}

\author{Douglas P. Finkbeiner$^{(a)}$, Tracy R. Slatyer$^{(b)}$, Neal Weiner$^{(c)}$, and Itay Yavin$^{(d)}$ \\ {\it (a)  Harvard-Smithsonian Center for Astrophysics, 60 Garden St., Cambridge, MA 02138, USA } \\ {\it (b) Physics Department, Harvard University, Cambridge, MA 02138, USA} \\ {\it (c) Center for Cosmology and Particle Physics, Department of Physics, New York University, New York, NY 10003, USA} \\ {\it (d) Department of Physics, Princeton University, Princeton, NJ 08544, USA}}

\abstract{Models of dark matter with $\sim$ GeV scale force mediators provide attractive explanations of many high energy anomalies, including PAMELA, ATIC, and the WMAP haze. At the same time, by exploiting the $\sim$ MeV scale excited states that are automatically present in such theories, these models naturally explain the DAMA/LIBRA and INTEGRAL signals through the inelastic dark matter (iDM) and exciting dark matter (XDM) scenarios, respectively. Interestingly, with only weak kinetic mixing to hypercharge to mediate decays, the lifetime of excited states with $\delta < 2 m_{\rm electron}$ is longer than the age of the universe. The fractional relic abundance of these excited states depends on the temperature of kinetic decoupling, but can be appreciable.  There could easily be other mechanisms for rapid decay, but the consequences of such long-lived states are intriguing.  We find that CDMS constrains the fractional relic population of $\sim 100$ keV states to be $\lsim 10^{-2}$, for a 1 TeV WIMP with $\sigma_n = 10^{-40}$ cm$^2$. Upcoming searches at CDMS, as well as xenon, silicon, and argon targets, can push this limit significantly lower. We also consider the possibility that the DAMA excitation occurs from a metastable state into the XDM state, which decays via $e^+e^-$ emission, which allows lighter states to explain the INTEGRAL signal due to the small kinetic energies required. Such models yield dramatic signals from down-scattering, with spectra peaking at high energies, sometimes as high as $\sim 1$ MeV, well outside the usual search windows. Such signals would be visible at future Ar and Si experiments, and may be visible at Ge and Xe experiments, although $\gamma$-rays associated with nuclear excitations would complicate the signal for these heavier targets. We also consider other XDM models involving $\sim 500$ keV metastable states, and find they can allow lighter WIMPs to explain INTEGRAL as well.}

\begin{document}
\section{Introduction}
\label{sec:intro}
The range of ideas for dark matter has recently expanded dramatically. Rather than limiting ourselves to candidates such as the MSSM neutralino, there has been a profusion of models with new forces, new annihilation modes, and multiple states. The motivation for these new properties has come from a variety of sources \footnote{See Ref. \cite{TODM} for a broader discussion.}, but dominantly from dramatic new astrophysical signatures, among them, PAMELA \cite{Adriani:2008zr}, WMAP \cite{2008arXiv0803.0732H}, ATIC \cite{Chang:2008zz} and PPB-BETS \cite{Torii:2008xu}.

Perhaps most striking among the recent signals is the PAMELA positron excess \cite{Adriani:2008zr}. The sharp rise of the positron fraction above 10 GeV seems inconsistent with interactions of CR protons with the interstellar medium, requiring a new primary source of positrons. This confirms the hints seen at HEAT \cite{Barwick:1997ig,Beatty:2004cy} and AMS-01 \cite{Aguilar:2007yf}. However, interpreting these signals as coming from dark matter annihilation is challenging in the traditional thermal WIMP scenarios. Annihilations to light leptons are often helicity suppressed \cite{Goldberg:1983nd}, while annihilation to hadronic or gauge boson modes provides spectra that are generally too soft to fit the data \cite{Cholis:2008hb,Cirelli:2008pk} \footnote{This could be alleviated if the signal is produced from a nearby clump of dark matter \cite{Hooper:2008kv}.}. Moreover, the copious anti-protons which are produced in these annihilations are generally an order of magnitude \cite{Cirelli:2008pk,Donato:2008jk} above the level seen by PAMELA \cite{Adriani:2008zq}. Finally, the cross section required to explain the positron excess is typically an order of magnitude (or more) larger than the thermal cross section of $3 \times 10^{-26} {\rm cm^3 s^{-1}}$.

A simple solution to all of these problems arises from postulating a new light force for dark matter \cite{TODM}. Annihilations of dark matter into light bosons which decay to electrons and muons can easily yield a hard lepton spectrum without significant anti-protons or $\pi^0$'s \cite{Cholis:2008vb} and provides a good fit to the PAMELA data \cite{Cholis:2008qq,Nelson:2008hj,Meade:2009rb,Mardon:2009rc}. Moreover, the light mediator can provide an enhanced cross section at low ($v \sim 220 {\rm km/s}$) velocities through the Sommerfeld effect \cite{TODM} or a capture into WIMPonium \cite{Pospelov:2008jd} \footnote{The Sommerfeld effect \cite{sommerfeld} was first discussed in the context of dark matter by \cite{Hisano:2004ds}. See also \cite{Cirelli:2007xd,MarchRussell:2008yu} for other recent discussions.}.

Such a model can also explains other astrophysical anomalies. Specifically, such annihilations can explain \cite{Cholis:2008vb,Cholis:2008wq} the WMAP ``haze'' \cite{Finkbeiner:2003im,Finkbeiner:2004us,Dobler:2007wv}, understood as high energy positrons and electrons from WIMP annihilation, synchrotron radiating in the galactic magnetic field \cite{Finkbeiner:2004us,Hooper:2007kb}. Excesses of electrons and/or positrons in the 400-700 GeV range reported by ATIC \cite{Chang:2008zz} and PPB-BETS \cite{Torii:2008xu} are also naturally explained \cite{Cholis:2008wq,Meade:2009rb,Mardon:2009rc} by the same annihilation channels leading to the positron excess in PAMELA.

If this new force is a gauge force, a very important consequence is the presence of quasi-degenerate states of dark matter \cite{TODM}. While such states can have important consequences for direct detection signals \cite{TuckerSmith:2001hy} or cosmic ray production \cite{Finkbeiner:2007kk}, it is important to recognize that such states are expected in these theories without any phenomenological input beyond PAMELA. Since Majorana fermions and real scalars cannot carry conserved charges, they cannot have diagonal couplings to gauge fields. As a consequence, there must be multiple Majorana fermions or real scalars in the theory, and the gauge couplings are off-diagonal. For the Sommerfeld effect to occur, these states {\em must} have splittings smaller than a typical WIMP kinetic energy, above which the enhancement is suppressed. Remarkably, with a GeV force carrier, the states are naturally generated from radiative effects in the $\sim$ 1 MeV range \cite{TODM}. Such splittings can lead to other signals with potentially observable ramifications. As we shall discuss, scatterings from the ground state into the excited state can provide simple explanations of the DAMA/LIBRA \cite{TuckerSmith:2001hy} and INTEGRAL \cite{Finkbeiner:2007kk} signals, using splittings of $\mathcal{O}(100~{\rm keV})$ and $\mathcal{O}({\rm MeV})$, respectively.

It has been previously noted \cite{Finkbeiner:2008gw} that excited states can have cosmologically interesting lifetimes and observable consequences, for instance on the CMB. These long-lived new states allow for a wide range of interesting phenomenology which we explore in this paper. In particular
\begin{itemize}
\item{Even with the large deexcitation cross sections that can arise with light mediators, it is possible to have significant relic populations of the excited states.}
\item{The lifetimes of these states are generically cosmologically long, although small couplings to the Z, or the presence of new, light states can shorten them dramatically.}
\item{There are strong direct detection constraints on the relic excited abundance from the unsuppressed down-scattering. If the abundance is near current limits, there may be a striking signal at upcoming Ge, Si and Ar experiments.}
\item{If the relic excited population is large, then the DAMA-associated splitting may arise from a metastable state transition into an excited state. That excited state then promptly decay into the true ground state via $e^+e^-$ emission. In this case, the ``exciting dark matter'' (XDM) proposal of \cite{Finkbeiner:2007kk} can accommodate lighter WIMPs. Such scenarios predict signals at direct detection experiments in the $800+$ keVr range.}
\end{itemize}
Again, we emphasize that such excited states are automatically present in any theory with a new gauge interaction, and that the splittings of these excited states {\em must} be at most $\mathcal{O}({\rm MeV})$, otherwise the Sommerfeld enhancement would be inoperative. As such, the motivation to consider inelastic up-scattering (as in \cite{Smith:2001hy}) or down-scattering from the excited state arises from the new theories related to PAMELA alone. Nonetheless, there are concrete reasons to consider specific scales for the excited states, associated with other possible dark matter anomalies, which we now review.

\subsection{DAMA and Inelastic Dark Matter}
The DAMA/LIBRA collaboration has recently reported a modulated signal at 8.2 $\sigma$ in their single-hit data, confirming their previous results from DAMA/NaI. The observed period and phase are that expected for WIMP dark matter. Unfortunately, no other experiment has confirmed this signal, and if the signal is from elastic nuclear scattering, recent limits from CDMS \cite{Akerib:2005kh,Ahmed:2008eu} and XENON10 \cite{Angle:2007uj} exclude this interpretation by over two orders of magnitude.

One proposal to reconcile these conflicting results is ``inelastic dark matter'' (iDM) \cite{TuckerSmith:2001hy}. In this model, dark matter transitions when scattering off of a nucleus into an excited state, with $m_{\chi^*} - m_\chi = \delta \sim 100 {\rm keV}$. The kinematic difference causes heavy targets to be favored over light targets, a shift of the spectrum to higher energies (suppressing or eliminating low energy events), and enhanced modulation \cite{TuckerSmith:2001hy,Tucker-Smith:2004jv}. Together, these weaken the bounds from other experiments sufficiently to allow the DAMA modulation to be consistent with the results of other experiments \cite{Chang:2008gd,MarchRussell:2008dy,Cui:2009xq}, even in light of the most recent results from XENON10, CDMS, ZEPLIN-III \cite{Lebedenko:2008gb,Lebedenko:2009xe}, CRESST \cite{Angloher:2008jj} and KIMS \cite{Lee.:2007qn}.

Models of iDM are simple to construct \cite{TuckerSmith:2001hy,Tucker-Smith:2004jv}. Pseudo-Dirac fermions as well as complex scalars, with slight non-degeneracies between real and imaginary components naturally give rise to inelastic transitions. Majorana fermions charged under a non-Abelian gauge group \cite{TODM} not only provide such phenomenology, but also explain the $\sim$ MeV scale splitting if the mediator is $\sim 100 {\rm MeV} - 1 {\rm GeV}$ \cite{TODM,ArkaniHamed:2008qp,Baumgart:2009tn}.

However, it is essential for these models that the excited state is largely unpopulated. Deexcitations (down-scattering) are visible at all experiments, and for $\mathcal{O}(1)$ populations of the excited state, would be significantly excluded. Smaller populations will provide for a remarkable signal at upcoming experiments. As we shall see in section \ref{sec:lifetime}, for the simplest models of iDM with a $\sim$ GeV mediator, the excited state is cosmologically stable. This motivates us to consider metastable versions of iDM, where the excited state can decay to $e^+e^-$.

\subsection{INTEGRAL and Exciting Dark Matter}
The INTEGRAL collaboration \cite{Weidenspointner:2006nu} has reported a significant signal of 511 keV radiation from the galactic bulge region, approximately gaussian with a FWHM of $6^\circ$, with an additional subdominant component correlated with the disk \cite{Weidenspointner:2007}. Interpreting the signal as arising from the capture and annihilation of positrons and electrons in parapositronium, this requires $3 \times 10^{42}\, e^+/{\rm sec}$.

The ``exciting dark matter'' (XDM) proposal of \cite{Finkbeiner:2007kk} postulated the existence of an excited state of dark matter $\chi^*$ with $m_{\chi^*}-m_\chi \sim {\rm MeV}$, which would decay via emission of $e^+e^-$. Such an idea takes advantage of the fact that the kinetic energy of a $\sim 500$ GeV WIMP in the galactic halo is approximately MeV. However, the large observed rate requires a large cross section as we now discuss.

The total rate of excited WIMP states in the galactic center is simply
\be
\tau^{-1} = \int_0^{rc}4 \pi r^2 dr \frac{\langle  \sigma_{ex} v \rangle}{2} \left( \frac{\rho(r)}{M_\chi} \right)^2,
\ee
where $\langle \sigma_{ex} v \rangle$ is the pairwise averaged excitation cross section, and $rc$ is the maximum radius included in the INTEGRAL signal (approximately 1 kpc). We can calculate this by using the Einasto profile \cite{Merritt:2005xc} with parameters set by the A-1 run of the Aquarius  simulation \cite{Springel:2008cc}, which is one of the highest resolution simulations to date (although, importantly, we should note it contains DM only). Doing so, one finds
\be
\tau^{-1} \approx 3 \times 10^{42} ~\frac{e^+}{\rm sec} \left( \frac{\langle{\sigma_{ex} v} \rangle}{2 \times 10^{-20}~{\rm cm^3 s^{-1}}} \right)\left(\frac{500~ {\rm GeV}}{M_\chi}\right)^{2}.
\ee
We have not accounted for uncertainties in the local density here, which could add an additional factor of a few to this estimate. This compares with the S-wave unitarity bound for a 500 GeV WIMP in the halo, moving at $v \sim 10^{-3} c$, of approximately $3 \times 10^{-19}\rm cm^{3} s^{-1}$.

Unfortunately, not all of the particles in the halo can participate, as many pairs are below threshold. Taking a 1-D RMS velocity of $v_{rms}= 200 ~{\rm km/s}$ one finds that approximately $1/20$ of the pairs are kinematically capable of scattering  \cite{Finkbeiner:2007kk}. Thus, with these parameters, only cross sections which approximately saturate the unitarity bound will yield a sufficiently large rate.

There are a few important comments to make here:  first, the presence of a light mediator for this process can naturally allow a near saturation of the unitarity bound, even if the couplings, themselves are pertubrative \cite{Finkbeiner:2007kk}. This arises simply from multiple particle exchanges. Second, the above fraction assumes Maxwell-Boltzmann velocity distribution with a {\em local} velocity dispersion of $v_{rms} =200~{\rm km/s}$, which is consistent with expectations arising from the galactic rotation. However, Maxwell-Boltzmann distributions underpredict the high-velocity component of the distributions when compared to simulations \cite{Springel:2008cc}. Moreover,  $v_{rms}$ could change as one moves to the galactic center, and the results are exponentially sensitive to this value. Determining the properties of the WIMP velocity dispersion is very difficult in the galactic interior, because the potential is dominated by baryons. In fact, recent simulations including baryons show a rise of $v_{rms}$ as one moves to the center of the galaxy \cite{Governato:2006cq,RomanoDiaz:2008wz,RomanoDiaz:2009yq}. For $v_{rms} \propto r^{-1/4}$, as argued in  \cite{RomanoDiaz:2008wz}, the velocity dispersion in the inner galaxy would be approximately twice as large, and the excitation process would be largely unsuppressed by the Boltzmann distribution. Hence, it is very plausible that even cross sections well below unitarity-satuation could explain the signal. Finally, recent measurements of the motion of the Milky Way \cite{Reid:2009nj} suggest that the rotation velocity is approximately 30 km/s higher than previously expected, additionally suggesting a mass 30\% larger than previously thought, lowering the needed cross section by a factor of 1.5 (from the density) and possibly significantly more (from the associated change in expectations for $v_{rms}$, which naturally scales with $v_{rot}$). Thus, while we believe the original XDM proposal remains viable, it is interesting to consider the scenarios which allow the WIMP to be lighter \footnote{Other authors \cite{Pospelov:2007xh}, using an NFW profile, and using a smaller $v_{rms}=180 {\rm km/s}$ conclude differently. We note that the arguments that the velocity dispersions in the inner galaxy go down based upon \cite{Ascasibar:2005rw} rely upon the assumption that the MW interior is dominated by dark matter, when it is, in fact, dominated by baryons. As we have noted, simulations involving baryons show the opposite trend  \cite{Governato:2006cq,RomanoDiaz:2008wz,RomanoDiaz:2009yq}.}.

\section{Lifetimes and relic populations of excited states}
\label{sec:popLT}
\label{sec:lifetime}

The thermal history of these dark matter models is somewhat different from a standard WIMP. Because of the new force carrier, which we denote with $\med$, the WIMP stays in thermal equilibrium via (a) $\chi \chi \leftrightarrow \med \med$ and (b) $\med e \leftrightarrow \gamma e$ \cite{Finkbeiner:2007kk,Pospelov:2007mp,Finkbeiner:2008gw,TODM,ArkaniHamed:2008qp}. Chemical freezeout occurs when the former interaction (a) becomes inefficient, which happens at the usual $T \sim m_\chi/20$. Kinetic decoupling from the SM bath occurs much later, however. The process $\chi \med \rightarrow \chi \med$ is efficient as long as  $\med$ remains relativistic, and in chemical equilibrium with the standard model. Chemical equilibrium is maintained until $T \sim m_\med$, because the reaction (b) is very efficient, with cross section $\sigma \sim \alpha^2 \epsilon^2 m_\med^{-2}$ \cite{Finkbeiner:2008gw}. As a consequence, kinetic decoupling can occur as late as $T \lsim 30-300 {\rm MeV}$, or even later for lighter force mediators. After this point, the temperature of the dark sector scales as $a^{-2}$. Consequently, the DM can reach $T \sim 100~{\rm keV}$ before SM BBN occurs.

In theories where the new force in the dark sector is identified with a spontaneously broken gauge symmetry, as in Refs. \cite{TODM,ArkaniHamed:2008qp}, the situation has an added element. New degrees of freedom associated with the higgsing of the gauge symmetry are expected and can have long lifetimes themselves \cite{TODM,ArkaniHamed:2008qp,Baumgart:2009tn,Katz:2009qq,Cheung:2009qd}. For example, a higgsino/gaugino-like state can have a long lifetime owing to its suppressed decay into a gravitino. Depending on the precise spectrum and nature of these states their relic abundance can span several orders of magnitude $\Omega_h \sim .1 - 10^5$. If such particles have lifetimes $\tau \lsim 10^4 {\rm sec}$, as they decay electromagnetically, their energy is harmlessly deposited into the photon bath \cite{Jedamzik:2006xz}.  However, in such a situation, the number density of particles off of which $\chi^*$ can down-scatter is many orders of magnitude higher. Therefore, the effects of these light scatterers on the population of the excited WIMP state can be substantial.

\subsection{The Relic Population of Excited WIMP states}
In the early universe, the populations of excited and ground states are equal, but as the universe cools, the number of excited states becomes Boltzmann suppressed, so long as deexcitation scatterings remain active. Indeed, for large (MeV) splittings, it is quite natural to have a small, but non-zero, relic fraction \cite{Finkbeiner:2008gw}.

The degree to which the excited states remain present depends on a number of factors. First is simply the number of particles off of which an excited WIMP can collisionally deexcite. In some cases, $\chi^*$ can only deexcite via $\chi^* \chi^* \rightarrow \chi \chi$, in which case the number of scatterers falls off exponentially. In other cases, for instance with scalar or other interactions, $\chi^* \chi \rightarrow \chi \chi$ is also possible, and the rate for down-scattering is higher since the scatterer, $\chi$, maintains an unsuppressed number density. Finally, the presence of long-lived ($\tau \sim 1~{\rm sec}$) light states, with possibly large population, offers additional scatterers against which $\chi^*$ can deexcite.  

A second important factor that determines the fractional abundance of excited states is the decoupling temperature of the WIMPs \cite{Finkbeiner:2008gw}. In particular, if the WIMPs kinetically decouple earlier, then the resulting relic fraction of excited states tends to be lower. This is so because after kinetic decoupling the temperature of the dark sector drops more rapidly than the photon temperature and therefore the deexcitation reaction remains in equilibrium longer. The upshot of all this, as we shall see, is that there is a wide range of possible excited fractions, from $\mathcal{O}(1)$ to negligibly small.

To quantify this, we solve the Boltzmann equation in the limit where $d n/dt = 0$ to obtain the temperature at freeze-out and hence the relic abundance. 
\begin{equation}
3 H(z) = \langle \sigma_{ex} v \rangle n_{S}(z) 
\end{equation}
where $H(z)$ is the Hubble parameter as a function of redshift, and $n_{S}(z)$ is the number density of downscattering targets. For self-scattering $n_S = n_{\chi^*}= n_\chi e^{-\delta/T_W}$ where $T_W$ is the temperature of the WIMPs (which can, in general, be different from the photon temperature). For more general interactions $n_S = n_\chi$, and for scatterings off of other long-lived particles X $n_S = n_X$. The fraction of excited states at freeze-out is then essentially just $\exp(-\delta/T_{Wf})$.

The relic abundance of the excited population drops as the WIMP kinetic decoupling occurs earlier \cite{Finkbeiner:2008gw} (because the WIMP temperature scales as $T = T^2_\gamma/T_{kd}$, where $T_{kd}$ is the temperature of kinetic decoupling). We thus consider two separate cases: first, we consider the case where $T_W = T_\gamma$, which should be thought of as a limiting case with the maximal relic excitation fraction. Because kinetic equilibrium is maintained generally by the thermal (chemical) equilibrium of the mediator $\med$, this typically will not occur for splittings $\delta \lsim {\rm MeV}$. Alternatively, we consider $T_{kd} = 1~{\rm GeV}$, which is a reasonable, if somewhat high value for $T_{kd}$.

We further distinguish between the possibility that $\chi^* \chi^* \rightarrow \chi \chi$ is the only deexcitation process, against the situation where $\chi^* \chi \rightarrow \chi \chi$ also contributes. In the former case, deexcitation terminates earlier as the excitated fraction drops exponentially.

\begin{figure}
\begin{center}
\includegraphics[scale=.8]{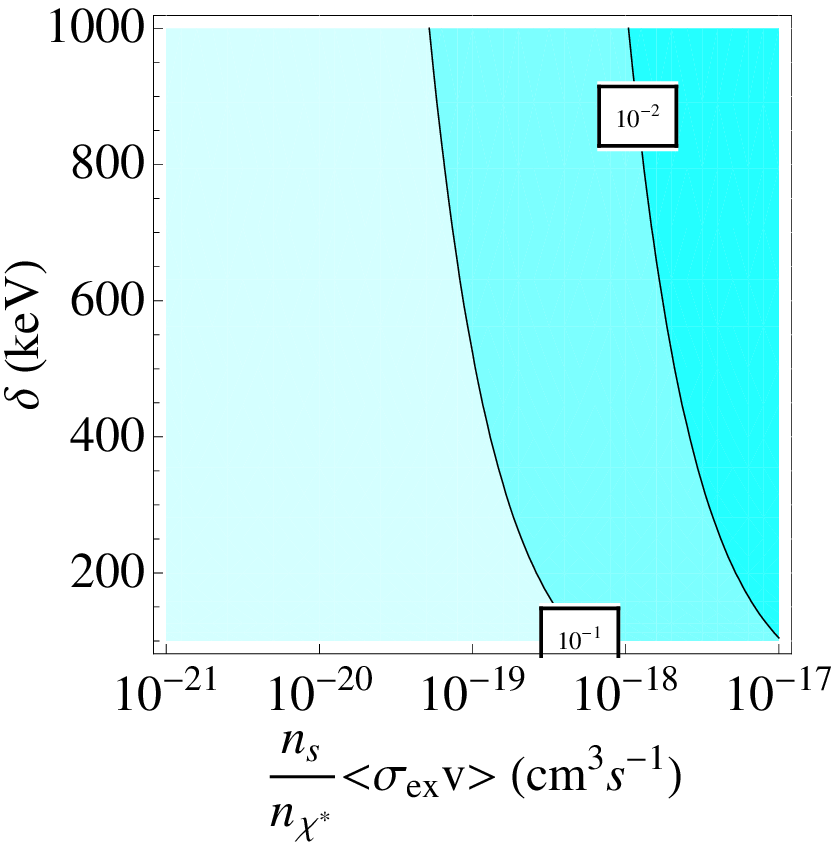}
\includegraphics[scale=.8]{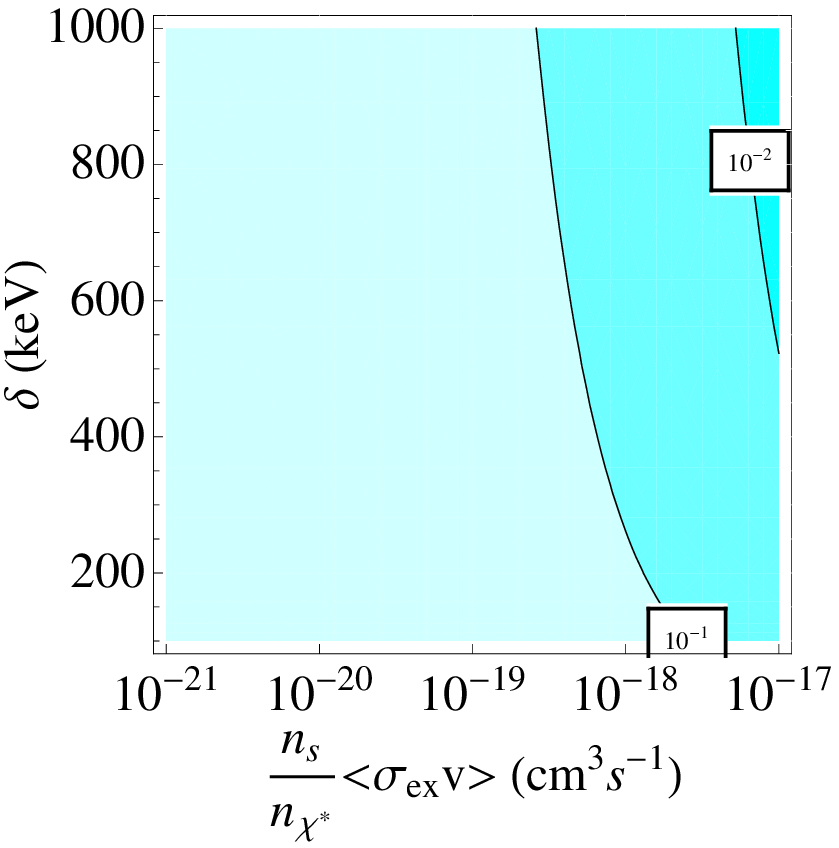}
\end{center}
\caption{Contour plots for the fractional abundance of the excited state after freeze-out, plotted against $n_s/n_{\chi^*} \langle \sigma_{ex} v \rangle$ (assuming the number of scatterers $n_s$ is exponentially suppressed by $e^{-\delta/T}$), and the excitation gap, $\delta$. We consider a WIMP mass of $100$ GeV ($500$ GeV) on the left (right).}
\label{fig:FractionExp}
\end{figure}

In Fig. \ref{fig:FractionExp} we plot the fractional abundance as a function of the thermal excitation cross-section $\langle \sigma_{ex} v \rangle$ and the excitation energy, $\delta$ for the case where $\chi^* \chi^* \rightarrow \chi \chi$ acts alone and  $T_W = T_\gamma$. As can be seen from the plots, this fraction is $\mathcal{O}(1)$, unless we consider very large thermal cross-sections, $\langle \sigma_{ex} v \rangle \gtrsim 10^{-19} {\rm cm^3 s^{-1}} $ or very large excitation energies $\delta \gg \mathrm{MeV}$. Notice that the fractional abundance is lower for a lighter WIMP. This is a consequence of the fact that a larger number density leads to a higher reaction rate and hence to a lower freeze-out temperature and further depletion. 

\begin{figure}
\begin{center}
\includegraphics[scale=.76]{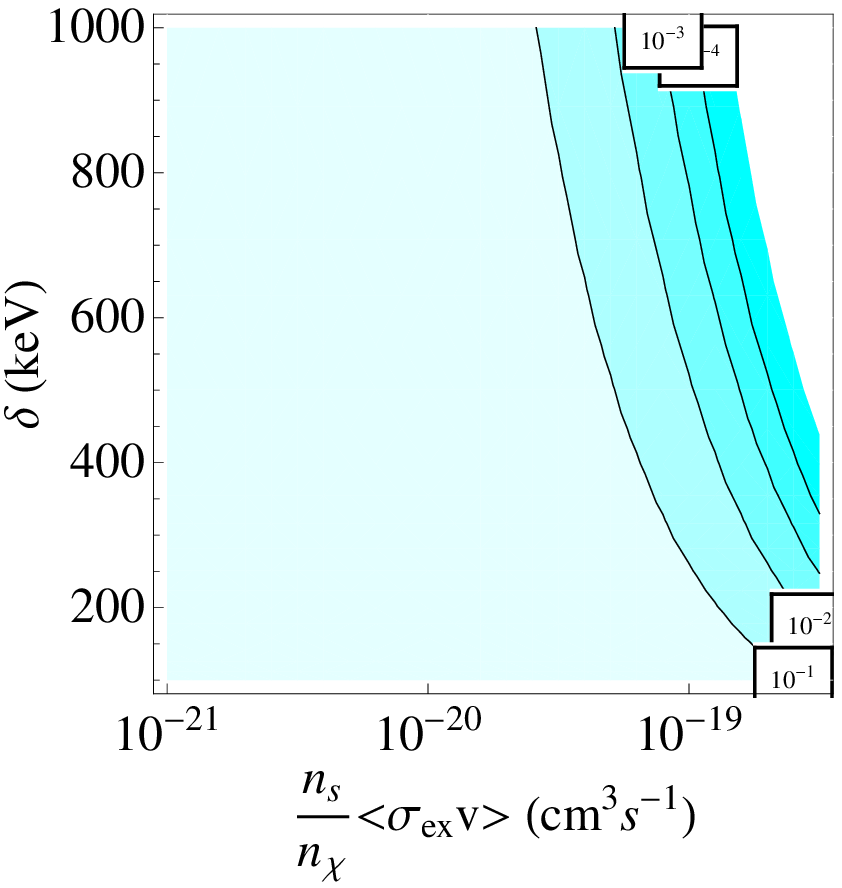}
\includegraphics[scale=.8]{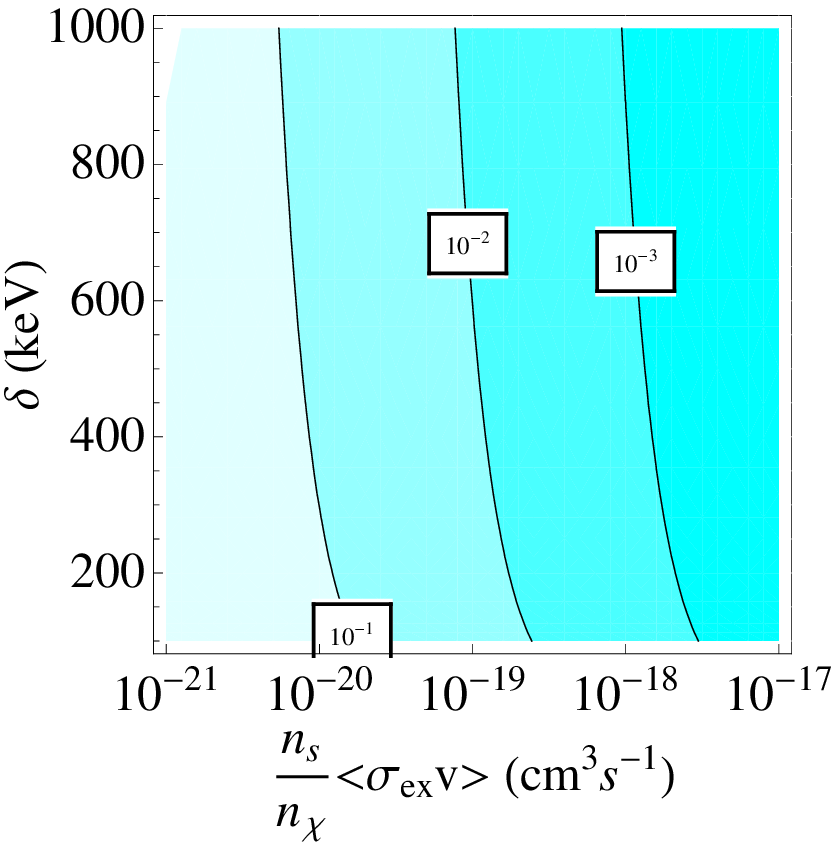}
\end{center}
\caption{On the left is a contour plot for the fractional abundance of the excited state after freeze-out, plotted against $n_s/n_\chi \langle \sigma_{ex} v \rangle$ (assuming the number of scatterers $n_s$ is \emph{not} exponentially suppressed by $e^{-\delta/T}$), and the excitation gap, $\delta$. On the right we assume $n_s$ is exponentially suppressed by $e^{-\delta/T}$ as in Fig. \ref{fig:FractionExp}, but with kinetic decoupling at $T_{kd} = 1$ GeV. In both cases the WIMP mass is fixed at $500$ GeV.}
\label{fig:FractionAlt}
\end{figure}

In contrast, in Fig. \ref{fig:FractionAlt}, we depict two alternative scenarios and the resulting fraction. On the left we consider a situation where $\chi^* \chi \rightarrow \chi \chi$ is present as well and therefore the density of scatterers does not deplete exponentially. On the right, we consider a situation where the decoupling temperature is $T_{dec} = 1$ GeV. Both effects tend to wash out the excited state population. If the two effects are combined, they may result in negligible abundance. 

If there are any light states in the spectrum which are long-lived and can participate in the deexcitation process, then depletion of the excited state can be even more complete. Since their number density can be considerably higher than that of DM, the light scatterers lead to very efficient depopulation of the excited state. As can be seen on the left pane of Fig. \ref{fig:FractionAlt}, when $n_s/n_\chi \gg 1$ the resulting fractional abundance can be extremely small. In section \ref{sec:models} we give a simple example which realizes such a scenario. 

In light of the above, it is clear that the fractional abundance of the excited state can range over many orders of magnitude and depends on the interactions and particle content present in the dark sector. It is therefore important to consider the different alternatives and their respective signatures. If a large population of the excited state is present, then bounds from current direct detection experiment place restrictions on the size of the deexcitation cross-section and/or the deexcitation energy. It also calls for current experiments to search for such deexcitation events as we discuss in section \ref{sec:signals}. Finally, in the models where the iDM and XDM transitions occur from a metastable state with a sizeable relic fraction, as we discuss below,  the XDM scenario to explain  INTEGRAL is easier to realize since the energy threshold is considerably lower.
   
\subsection{The Lifetime of Excited States}
If the excited state can only relax down through a gauge-boson in its sector which is kinetically mixed with the SM, then the lifetime of such a state is longer than the age of the universe, unless the energy gap $|\delta| > 2m_{\mathrm{electron}}$. Above the electroweak scale the only mixing term allowed is, 
\begin{equation}
\mathcal{L} \supset -\frac{\epsilon}{2} \med_{\mu\nu} B^{\mu\nu}
\end{equation}
where $B_{\mu\nu}$ ($\med_{\mu\nu}$) is the hypercharge (dark abelian) field strength. The mixing parameter, $\epsilon$, is constrained to be $\epsilon \lesssim 10^{-3}$ \cite{Pospelov:2008zw}. This operator, below the electroweak scale, leads to mixing with the photon as expounded on in Refs.\cite{Pospelov:2007mp,TODM,ArkaniHamed:2008qp}. It also results in mixing of the dark sector with the $Z^0$ vector-boson which now couples to the dark abelian current. Below the $Z^0$ mass we should integrate it out to generate a coupling between the abelian gauge-boson and the neutrino current. That coupling will render the excited state unstable, and allow it to decay through an off-shell emission of two neutrinos. The lifetime for that process is given by
\be
\Gamma_{\nu\nu} &\approx& \Gamma_n \left(\frac{\ald}{\alpha_2} \right) \left(\frac{\epsilon M_Z^2}{m_{\med}^2} \right)^2\left(\frac{|\delta|}{m_n - m_p} \right)^5 \left(\frac{|\delta|}{M_Z} \right)^4 \\\nonumber &=& \frac{1}{3\times 10^{21}~\mathrm{sec}} \left(\frac{\delta}{\mathrm{MeV}} \right)^9\left(\frac{\epsilon}{10^{-3}} \right)^2,
\ee
where $m_n$, $m_p$, $M_Z$, and $m_{\med}$ is the mass of the neutron, proton, $Z^0$ vector-boson, and dark vector-boson, respectively. $\Gamma_n^{-1}$ is the neutron lifetime, and we have taken the dark gauge coupling to be $\ald = 1/137$. 

A faster decay mode is for the excited state to decay through an off-shell emission of 3 photons. Similar to the Euler-Heisenberg Lagrangian, below the electron mass one generates a gauge-invariant interaction between the dark gauge-boson and 3 photons,
\begin{equation}
\mathcal{L}\supset \frac{\epsilon \cos\theta_W}{90~ m_e^4}\left( \med_{\mu\nu}F^{\mu\nu} F_{\alpha\beta}F^{\alpha\beta} +\frac{4}{7} \tilde{\med}_{\mu\nu}F^{\mu\nu}\tilde{F}_{\alpha\beta}F^{\alpha\beta}\right)
\end{equation}
where the tilde denotes the dual field and $\theta_W$ is the Weinberg angle. We estimate the lifetime by scaling the neutron lifetime and adding an extra $1/16\pi^2$ to account for the additional phase-space suppression,
\be
\label{eqn:3photon}
\Gamma_{3\gamma} &\approx& \Gamma_{n} \times \left(\frac{\alpha\ald}{\alpha_2^2}\right) \times \left(\frac{\theta\epsilon M_Z^2 }{m_{\med}^2}\right)^2 \times \left( \frac{\alpha^3}{4\pi}\right)\times \left(\frac{|\tilde \delta|}{m_n-m_p}\right)^5\times \left(\frac{|\tilde \delta|^4}{90~ m_e^4}\right)^2 \\\nonumber \\\nonumber&=&\frac{\theta^2}{8\times 10^{19}~{\rm sec}}\left(\frac{\tilde \delta}{\mathrm{300 keV}}\right)^{13} \left(\frac{\epsilon}{10^{-3}} \right)^2
\ee
where $\tilde \delta$ is the effective available phase space (which is shared between three photons), and $\theta$ parametrizes any additional suppression present in the coupling beyond that introduced by the kinetic mixing coefficient, $\epsilon$. In the models presented below $\theta$ can be naturally small in which case the resulting lifetime is extremely long. The combination $\theta\epsilon/m_{\med}^2$ is bounded from above by direct detection experiments. The precise value depends strongly on the deexcitation energy $\delta$, but for $\delta \sim 100$ keV it is  $\theta\epsilon/m_{\med}^2 < 10^{-3}/M_Z^2$ (see Fig. \ref{fig:deltaVssigma200} below). 

If the excitation of the metastable state to the excited state in the GC is to explain the INTEGRAL signal via the XDM mechanism, the excited state must relax to the ground state before it traverses a distance larger than $\sim$ kpc. This places a constraint on its lifetime,
\begin{equation}
\label{eqn:kpc}
v \tau^* \sim 1~\mathrm{kpc}\quad \quad \tau^* < 10^{14}\mathrm{sec}
\end{equation} 
where $\tau^*$ is the excited state's lifetime and we assumed $v\sim 10^{-3}$. The excited state relaxes to an electron-positron pair via an off-shell dark gauge-boson emission, and its lifetime is almost identical to that of the neutron,
\begin{equation}
\label{eqn:XDMdecay}
\Gamma = \frac{1}{\tau^*} = \Gamma_n \times \left(\frac{\alpha\ald}{\alpha_2^2}\right) \times \left(\frac{\theta\epsilon M_Z^2 }{m_{\med}^2}\right)^2 \left(\frac{2m_e}{m_n-m_p}\right)^5 = \frac{\theta^2}{900~\mathrm{sec}} \left(\frac{\epsilon}{10^{-3}} \right)^2
\end{equation}
We again use $\theta$ to parametrize any additional suppression in the excited to ground state transition. Using Eq. (\ref{eqn:kpc}) we can place a lower bound $\theta \gtrsim  10^{-5}$ from requiring the transition to happen within a kpc of the GC. We note that the $\theta$'s utilized in Eqs. (\ref{eqn:3photon}) and (\ref{eqn:XDMdecay}) are logically disjoint as they pertain to different transitions. 

\subsection{Other Decay Mechanisms for Excited States}
Before proceeding, it is important to consider the viable possibilities for a more rapid decay of the excited state. Essentially, the appearance of {\em any} light states with appreciable couplings to the WIMPs can lead to rapid decay. A few options are neutrinos (through direct Z-coupling), axions, or right-handed neutrinos coupled to light states in the dark sector.

A very simple possibility is to consider the presence of some state with hypercharge (something akin to a Higgsino, for instance), which the dark matter particle mixes with by an amount $\xi$. The excited state in this case can decay via an offshell Z boson to $\nu \bar \nu$. Even for $\xi \lsim 10^{-5}$ and $\delta \sim 100~{\rm keV}$, the particle will decay quickly on cosmological timescales. 

Another possibility is decays to axions. Although they are not the principle dark matter candidate in these models, they still serve the important purpose of solving the strong CP problem. If the splitting between excited and ground states occurs radiatively (or some component of it does), then there can be a decay into axions with lifetimes
\be
\Gamma_a \sim \frac{1}{32 \pi} \left( \frac{\delta_a}{f_a} \right)^2 \delta
\ee
where $\delta_a$ is the contribution to the splitting which occurs spontaneously, and $f_a$ is the axion decay constant. Even for $\delta_a \sim 1$ keV, the lifetime is cosmologically short.

Finally, we can consider the possibility of new, light states in the dark sector that couple to right handed neutrinos. The dynamics in the dark sector leading to vector-boson masses of $\sim$ GeV often involves some light scalar fields which could couple to the DM states as well. A neutral state in the spectrum (such as discussed in \cite{Katz:2009qq,Cheung:2009qd}) may also couple to right-handed neutrinos directly. Mixing between the different light states leads to an effective coupling between the DM states and right-handed neutrinos. If right-handed neutrinos are sufficiently light, then the DM excited state can relax into them on a cosmologically short timescale. 

\section{Scenarios and Models for metastable iDM/XDM}
\label{sec:models}
One alternative to having inelastic dark matter excite from the ground state into a $100~$keV state is to imagine that there is a large population of WIMPs in a metastable state just below the $e^+e^-$ threshold of 1022 keV. In this case, the inelastic transition would be into a state that could efficiently decay into the true ground state \footnote{This same spectrum was recently employed by \cite{Chen:2009dm}.}. Such a model would also allow the XDM scenario to be viable for lighter ($\sim 200$ GeV) WIMPs. Although these particles would not explain the ATIC/PPB-BETS excesses, they would still be able to naturally explain PAMELA and the Haze. It is unclear what the limits on deexcitations from a $\sim 800-900$ keV state would be, as they are outside the search windows of CDMS and XENON. An examination of these data could likely exclude a deexcitation rate with similar cross section to that for the iDM excitation. Still, because they arise from different gauge bosons, with {\em a priori} different mixings with the photon, the cross sections can differ by many orders of magnitude, as we shall show.

To demonstrate this phenomenology, we consider here some simple field theories that result in a spectrum with a metastable state and an ample WIMP-WIMP inelastic cross section, but with suppressed deexcitation cross section. These models easily arise when the dark sector gauge bosons are not equally mixed with the SM photon. The first example we will provide illustrates the possibility of a metastable state with the necessary couplings to scatter inelastically off nuclei to an excited state which is $\sim100$ keV above it as in the iDM scenario. The excited state then rapidly relaxes to the ground state through the same transition responsible for the INTEGRAL signal (Fig. \ref{fig:su2GenSplit}).

Although it is incompatible with the iDM scenario, we will also provide an intriguing second example. Here, we will exhibit a scenario with a metastable state which can scatter both up and down through the same gauge-boson. This allows for annihilations of the metastable state into the excited state which is unsuppressed by kinematics. As a result, the annihilations in the center of the galaxy which lead to the INTEGRAL signal are not restricted to a small fraction of the WIMP distribution, but rather occur between any pair of metastable states.

Finally, we briefly discuss models with light scatterers that can deplete the excited state's population as argued in section \ref{sec:popLT}. 

\subsection{Radiative Splittings and the iDM Scenario}
\label{subsec:inverted}
We begin by addressing a scenario in which both DAMA and INTEGRAL are explained by a $\sim 100~{\rm keV}$ inelastic scattering from a metastable state ($\delta \approx 900~{\rm keV}$) into a more rapidly decaying excited state ($\delta \approx 1~{\rm MeV}$). The downscattering process is generally suppressed, because the different gauge bosons naturally mix with different strengths to hypercharge.

We begin with a model with $SU(2)$ gauge symmetry under which DM transform as a triplet. The Lagrangian also contains two scalar adjoints of $SU(2)$, $\phi$ and $\phi^\prime$,
\be
\label{eqn:SU2Lag}
\mathcal{L} &\supset&  \frac{1}{2} M\chi \chi +  V(\phi,\phi^\prime) \\\nonumber
&+& \frac{1}{\Lambda} \tr\left(\phi w^{\mu\nu}\right) B_{\mu\nu} +  \frac{1}{\Lambda} \tr\left(\phi^\prime w^{\mu\nu}\right) B_{\mu\nu} + \frac{1}{\Lambda^2}\left( \phi_i \phi^\prime_j w^{\mu\nu}_k \epsilon^{abc}\right) B_{\mu\nu}
\ee
where $w^{\mu\nu}$ ($B^{\mu\nu}$) is the dark SU(2) (SM hypercharge) field strength, and $\Lambda \sim~\mathrm{TeV}$ is some high energy threshold.

\begin{figure}
\begin{center}
\includegraphics[scale=.7]{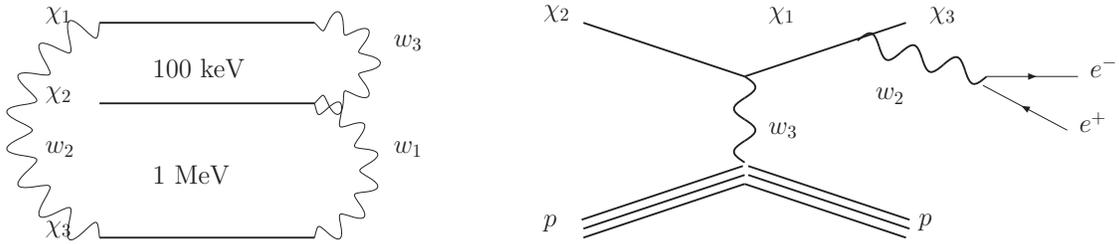}
\end{center}
\caption{On the left pane we show the splittings induced in the $SU(2)$ triplet by radiative corrections, as well as all the relevant couplings. The mixing of the $w_1$ gauge boson to the SM hypercharge is further suppressed compared with the mixing of $w_2$ and $w_3$. This leads to the possibility of deexcitation processes that can be seen in direct detection experiments. The inelastic scattering off nuclei is depicted on the right. }
\label{fig:su2GenSplit}
\end{figure}

We assume that the scalar potential, $V(\phi,\phi^\prime)$ is such that $\phi$ and $\phi^\prime$ get a VEV and $\langle\phi_3\rangle \ne 0$ and $ \langle \phi^\prime_2\rangle\ne 0$. The gauge symmetry is broken completely and all 3 gauge-bosons become massive. At low energies, the fermions receive a radiative correction that splits their masses according to,
\begin{eqnarray}
\nonumber
\delta M_1 &=& \frac{\gd \ald}{2} \left( \langle \phi_3 \rangle +  \langle \phi^\prime_2 \rangle \right) \\
\nonumber
\delta M_2 &=& \frac{\gd \ald}{2} \langle \phi_3 \rangle\\
\delta M_3 &=& \frac{\gd \ald}{2} \langle \phi^\prime_2 \rangle
\end{eqnarray}
where $\gd$ is the dark $SU(2)$ gauge coupling, $\ald = \gd^2/4\pi$, and we ignored corrections of order $\langle\phi\rangle/M$. In the limit where $\langle \phi^\prime_2\rangle = 0$ (or  $\langle \phi_3\rangle = 0$)  we restore an unbroken $U(1)$ with $\chi_3$ ($\chi_2$) a neutral state which does not receive a mass correction. With some tuning, it is possible to achieve a spectrum (Fig. \ref{fig:su2GenSplit}, left panel) which contains all the appropriate scales for iDM and XDM.

The higher dimensional operators in the Lagrangian of Eq. (\ref{eqn:SU2Lag}) will result in mixing of $w_3$ and $w_2$ with the SM hypercharge. The mixing parameter is order $\langle\phi\rangle/\Lambda \sim 10^{-3}-10^{-4}$ if the gauge bosons are at $\sim \mathrm{GeV}$. As shown on the right of Fig. \ref{fig:su2GenSplit} this mixing generates the transitions necessary to explain both DAMA and INTEGRAL in the iDM and XDM scenarios, respectively. 

At higher order, $w_1$ is also mixed with hypercharge at order $\langle\phi\rangle\langle\phi^\prime\rangle/\Lambda^2 \sim 10^{-6}-10^{-8}$. This coupling will allow for deexcitation transitions in direct detection experiments which we discuss in the next section. Note that in this case, $\theta = \langle\phi^\prime\rangle/\Lambda \lesssim 10^{-3}$ and the lifetime in Eq. (\ref{eqn:3photon}) is considerably longer than the age of the universe. 

\subsection{Up-Down Transition for INTEGRAL}
In view of the possible significance of metastable states, another intriguing possibility arises, where metastable states scatter $\chi_1 \chi_1 \rightarrow \chi_2 \chi_0$. This can yield a kinematically unsuppressed signal for INTEGRAL, although it does not provide us an obvious interpretation for DAMA.
We consider again an $SU(2)$ model, but take the DM fermions to be a massive Dirac triplet while the scalar sector consists of two real triplets, $\phi$, $\phi^\prime$ as before,
\begin{eqnarray}
\label{eqn:up-downLag}
\mathcal{L} &\supset& M \chi\chi^c + y \chi_i \phi_j \chi^c_k \epsilon^{ijk}  + V\left(\phi,\phi^\prime\right) \\\nonumber
&+& \frac{1}{\Lambda}\tr\left(\phi w^{\mu\nu}\right) B_{\mu\nu} + \frac{1}{\Lambda} \tr\left([\phi^\prime, \chi] [\phi^\prime, \chi] \right) 
\end{eqnarray}

\begin{figure}
\begin{center}
\includegraphics[scale=.7]{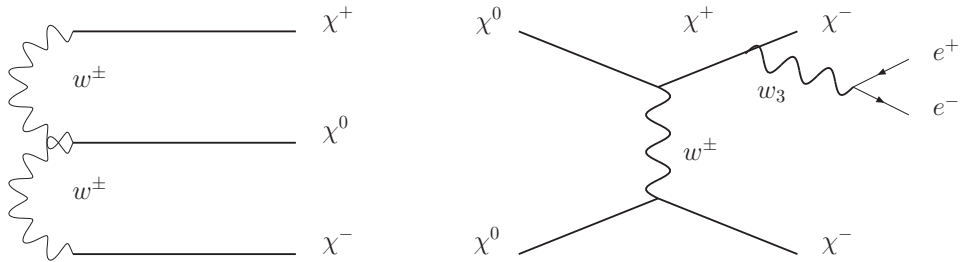}
\end{center}
\caption{The spectrum resulting from the Lagrangian in Eq. (\ref{eqn:up-downLag}) is shown on the left pane together with the relevant couplings. The up-down scattering associated with INTEGRAL is shown on the right. Notice that dark charge is softly broken by the condensation of $\phi^\prime$, hence allowing the seemingly charge violating transitions.}
\label{fig:up-downSplit}
\end{figure}

When $\phi$ condenses, $\langle\phi_3\rangle \ne 0$, it breaks $SU(2)$ down to $U(1)$ and renders the appropriate gauge-bosons massive. It also generates a splitting in the fermionic sector between the neutral and charged components (charged with respect to the unbroken $U(1)$), as shown on the left in Fig. \ref{fig:up-downSplit}. Notice that the Yukawa coupling needs to be fairly small to allow for the MeV splittings needed for INTEGRAL. 

At this point, the model contains dangerous elastic transitions as well as a long range force ($w_3$ is still massless). Also, the excited state $\chi^+$ cannot decay down to $\chi^-$ and generate the electron-positron pair needed for INTEGRAL. The condensation of $\phi^\prime$ resolves these problems. It splits the Dirac pairs $\chi^\pm$ into Majorana components with a mass difference of $\langle \phi^\prime_2\rangle^2/\Lambda$ and suppresses the elastic transition. It also softly mixes $\chi^+$ with $\chi^-$ by an amount $\langle \phi^\prime_2\rangle^2/(y\langle \phi_3\rangle \Lambda)$ to allow for a decay through $w_3$ (which both states couple to independently).
 
\subsection{Models with Additional Light Scatterers}
As we saw above, a wide range of models is possible. If the population of excited states is significant, we can have the INTEGRAL and iDM transitions occur from a metastable state (i.e., the inverted spectrum of subsection \ref{subsec:inverted}), or from pairwise up-down scattering. However, if we employ a spectrum such as considered in Ref. \cite{Smith:2001hy} or \cite{TODM}, where the lightest two states are split from one another by $\sim 100$ keV, the relic population of the excited states may be non-negligible, and it is important to consider the full range of how many of these states are present in the universe today.

The reason for this is simple: as we shall describe, deexcitation scatterings (i.e., exothermic transitions) are capable of occuring at {\em any} direct detection experiment. While transitions from $\sim 300+$ keV states are generally outside of the range studied by direct detection experiments, $\sim 100$ keV state down-scattering would be very visible, and is thus very constrained. As we shall see, with cross sections $\sigma_n \sim 10^{-40}{\rm cm^2}$, relic fractions larger than $\sim 10^{-2}$ are strongly constrained by CDMS.

Fractions of this size are roughly what is expected from deexcitations in the early universe through WIMP scattering. However, the presence of additional light states can diminish this fraction even further, by providing additional scatterers to keep the WIMP excited states in equilibrium until later times. Such states must satisfy three conditions: 1) they must be efficient scatterers; 2) they must live long enough to be present at the time of decoupling, but not too long as to spoil standard cosmological history; 3) they must have sufficiently large number density (compared with that of the DM states). 

The most obvious candidate to continue to depopulate WIMPs is the LSP of the new sector. For instance, the supersymmetric abelian models discussed in Refs. \cite{Katz:2009qq,Cheung:2009qd}  contain Higgsino-like states which mix with the gaugino associated with the abelian vector supermultiplet. In general, this results in mass eigenstates which are complete admixtures with elastic coupling to the dark force. Hence they form efficient scatterers if their lifetimes are long enough to allow them to be present at the time of decoupling. 

One possibility is that this LSP is simply stable, for instance in a gravity-mediated scenario, or a different state in the sector which is stable for a symmetry reason. Such a possibility of course requires that this particle is not the dominant component of the dark matter,  but that is easily the case, as we show below. Its number density can nevertheless be sufficiently high as it is much lighter than the DM states. 

On the other hand, LSP decay with a long lifetime is natural in gauge-mediated models. If the lightest fermion is heavier than any one of the dark gauge bosons then it can decay into it with the emission of a gravitino. The lifetime associated with this reaction is,
\begin{equation}
\label{eqn:binoLT}
\tau_{\tilde{\med}\rightarrow \med \tilde{G}} \sim 1{\rm~sec}\left(\frac{{\rm 300~MeV }}{m_{\med}} \right)^5 \left(\frac{\sqrt{F}}{100~{\rm TeV}} \right)^4 
\end{equation}
where $\tilde{\med}$ should be understood as the lightest fermion, which has some component of the lightest dark gauge boson $b$, and $F$ is the F-term vacuum expectation value that breaks supersymmetry. 

Finally, we must achieve a large number density of this light scatterer. If $\tilde{\med}$ is heavier than the lightest boson, then its relic abundance is determined by its annihilation into the lighter bosonic states. For example, in the zero-velocity limit, the annihilation into two dark gauge-bosons yields a cross-section \cite{Jungman:1995df},
\begin{equation}
\sigma(\tilde{\med}\tilde{\med} \rightarrow \med\med)v = \frac{\sin^4\phi \beta^3\gd^4}{8\pi m_{\tilde{\med}}^2}\left(\frac{1}{1-m_\med^2/m_{\tilde{\med}}} \right)^2
\end{equation}
where $\beta = (1-m_{\med}^2/m_{\tilde{\med}}^2)$, and $\sin\phi$ is the mixing between the LSP and the gaugino associated with the boson into which it is annihilating. In this case, $\Omega \lsim 1$ is natural (although could be much smaller if the only gauge boson in the dark sector is the one to which the DM dominantly couples).

On the other hand, if $\tilde{\med}$ is lighter than any of the bosons, its annihilation can only proceed into SM particles. Hence, it is suppressed by the kinetic mixing parameter and is given by,
 \be
 \sigma (\tilde{b}\tilde{\med}\rightarrow \mu^+\mu^-) v \sim \frac{\alpha^2 \epsilon^2}{m_{\med}^2} \left( \frac{m_\mu}{m_{\med}}\right)^2 \sim  \frac{\alpha^2}{m_Z^2} \left(\frac{m_{\med}^2}{m_Z^2}\right)
 \ee
 where we have used $\epsilon^2/m_{\med}^4 \sim 1/m_Z^4$ which is normalized to the DAMA signal, neglected the small suppression coming from the muon mass insertion, and taken $\ald \sim \alpha$.  Since the relic abundance of these particles is set by this cross section, we expect $\Omega \sim 10^2 - 10^5$.  In this case, there is some tension in the lifetime of the LSP, with decays into a SM photon and a gravitino the only channel kinematically allowed, which is further suppressed by $\epsilon^2$ compared to Eq. (\ref{eqn:binoLT}). Either a slightly higher mass or lower SUSY breaking scale would be required to secure a decay which is cosmologically safe.
 
While the vector-boson in this sector is very short lived and decays promptly into leptons, the scalar excitations may take longer to decay if they are lighter than the vector-boson (either through a four-body ($h\rightarrow 2e^+e^-$), or a loop-mediated two-body decay ($h \rightarrow e^+e^-$)). They can also satisfy the conditions of being efficient scatterers if additional Yukawa couplings with the DM states are present \footnote{The Yukawa coupling would have to be adequately small so not to introduce too large of splittings between the states. However, this small Yukawa could be compensated for by the very large number of Higgses.}.
 
In either case, significant populations of additional scatterers are present, which then further depopulate the excitated WIMPs. As a consequence, a wide range of excited fractions are possible, and thus a wide range of rates possible at upcoming experiments.
 
\section{Signals of excited states at direct detection experiments}
\label{sec:signals}
The presence of a relic population of excited states can have dramatic consequences for direct detection experiments. Because the down-scattering process $\chi^* N \rightarrow \chi N$ is unsuppressed by kinematics, this process is visible at any target, including light targets such as Ge and Ar, where inelastic up-scattering is highly suppressed \footnote{Similar down-scattering processes were considered in \cite{Bernabei:2008mv} with regards to light (sub-GeV) DM particles.}. Still, because the kinematics are different, the spectrum of events is peaked at high energies, similar to the up-scattering case, and thus motivates a broadening of the search window for all experiments.

We can categorize the interesting scenarios into a few specific cases:
\begin{itemize}
\item{The down-scattering can occur from a $\delta \sim 100 {\rm keV}$ state which is invoked to explain the DAMA signal from inelastic up-scattering. In this case, CDMS places strong constraints on the relic excitation fraction of the excited state, but strong signals are possible at future upgrades, as well as at Ar experiments.}
\item{The down-scattering can occur from a $\delta \sim 900~{\rm keV}$ state which is invoked to explain the INTEGRAL signal. In this case, the signal is outside the CDMS range of $10-100$ keVr. Because the energy transfer is so large, the simple form-factor approximation is not a good one, and many of the scatterings would be into excited nuclear states, which would decay promptly with photons. Such events would be vetoed at existing experiments, but it is possible that a small number of coherent scatterings would still occur at high energies.}
\item{Because the PAMELA/WMAP/ATIC signals compel us to consider a new force, which is naturally accompanied by splittings in the $\lsim$ MeV range, we are motivated a broad range of scales, generically. It is possible that other excited states in this range exist with significant relic densities, and their down-scattering signals might be unconstrained by current WIMP search energy ranges.}
\end{itemize}

\begin{figure}
\begin{center}
\includegraphics[scale=.53]{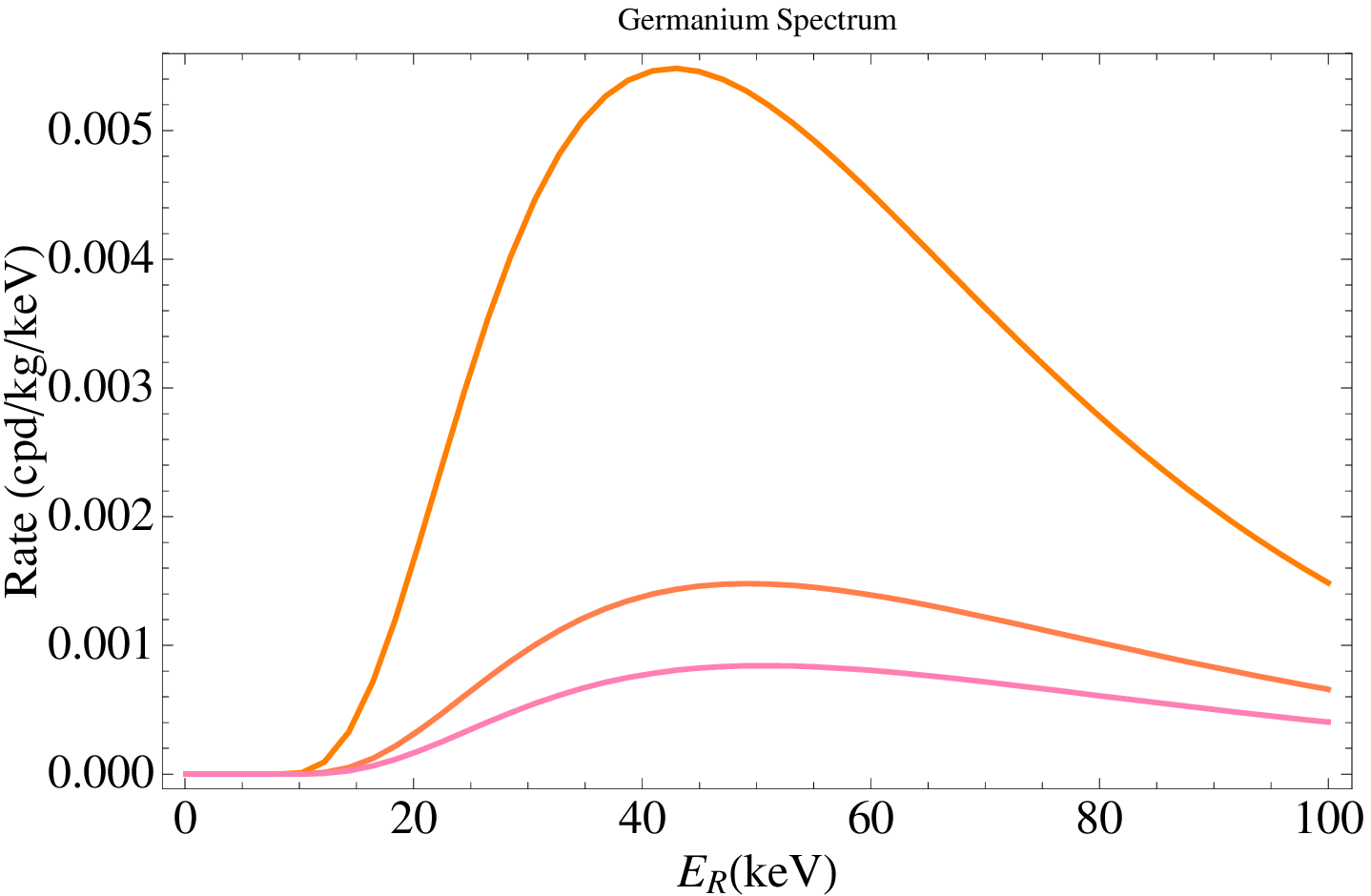}
\includegraphics[scale=.4]{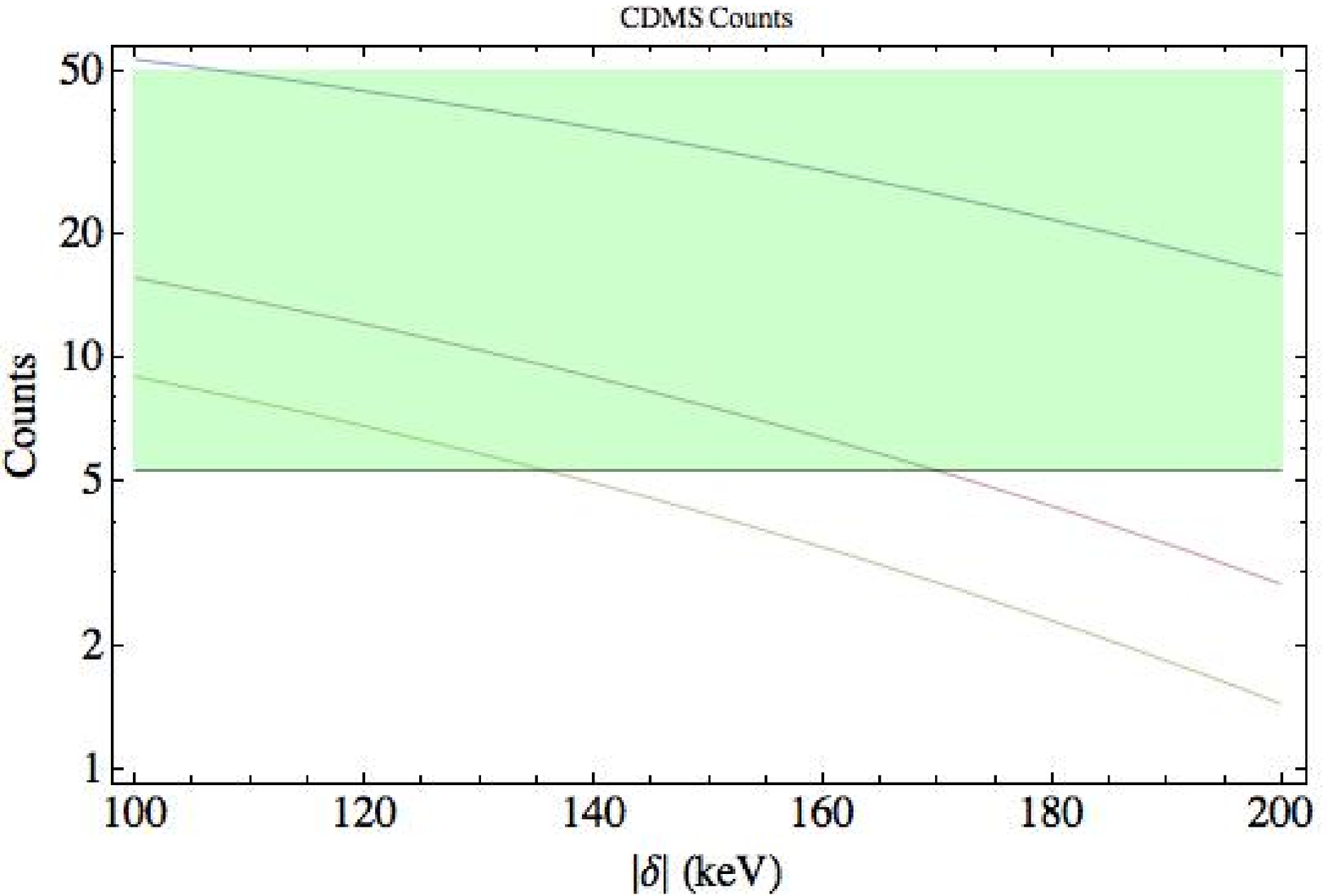}
\end{center}
\caption{In the left pane we depict the recoil energy spectrum expected in CDMS for a deexcitation transition with $|\delta| = 100$ keV. On the right we show the corresponding  predicted number of counts observed in CDMS as a function of the deexcitation energy gap, $|\delta|$. The WIMP-nucleon cross-section $\times$ fractional abundance was taken to be $F \sigma = 10^{-42}~\mathrm{cm}^2$. The WIMP mass is 100 GeV, 300 GeV, and 500 GeV, from top to bottom. The horizontal line at 5.3 counts marks the 90\% Poisson confidence upper limit on the expected number of signal events.}
\label{fig:CDMScounts}
\end{figure}

We calculate the scattering off of a nuclear target in the standard way \cite{Lewin:1995rx} recognizing the effects of inelasticity \cite{TuckerSmith:2001hy}. For a target of mass $\mN$ and an energy recoil $\ER$, the minimum WIMP velocity required to scatter is, 
\begin{equation}
 \beta_{min} = \frac{1}{\sqrt{2\mN \ER}}\left|\frac{\ER \mN}{\mu} + \delta \right| 
\end{equation}
where $\mu$ is the WIMP-nucleus reduced mass and $\delta < 0$ is the energy gap associated with the deexcitation. When $|\delta| \gg 100$ keV, scattering is dominated by events with very large energy recoil which are outside the experimental energy window. However, if $|\delta| \sim 100$ keV, then one must worry about existing bounds from CDMS as shown in Fig. \ref{fig:CDMScounts}. Such bounds should be properly interpreted as a limit on the product of cross section and relic excitation fraction. 

\begin{figure}
\begin{center}
\includegraphics[scale=.43]{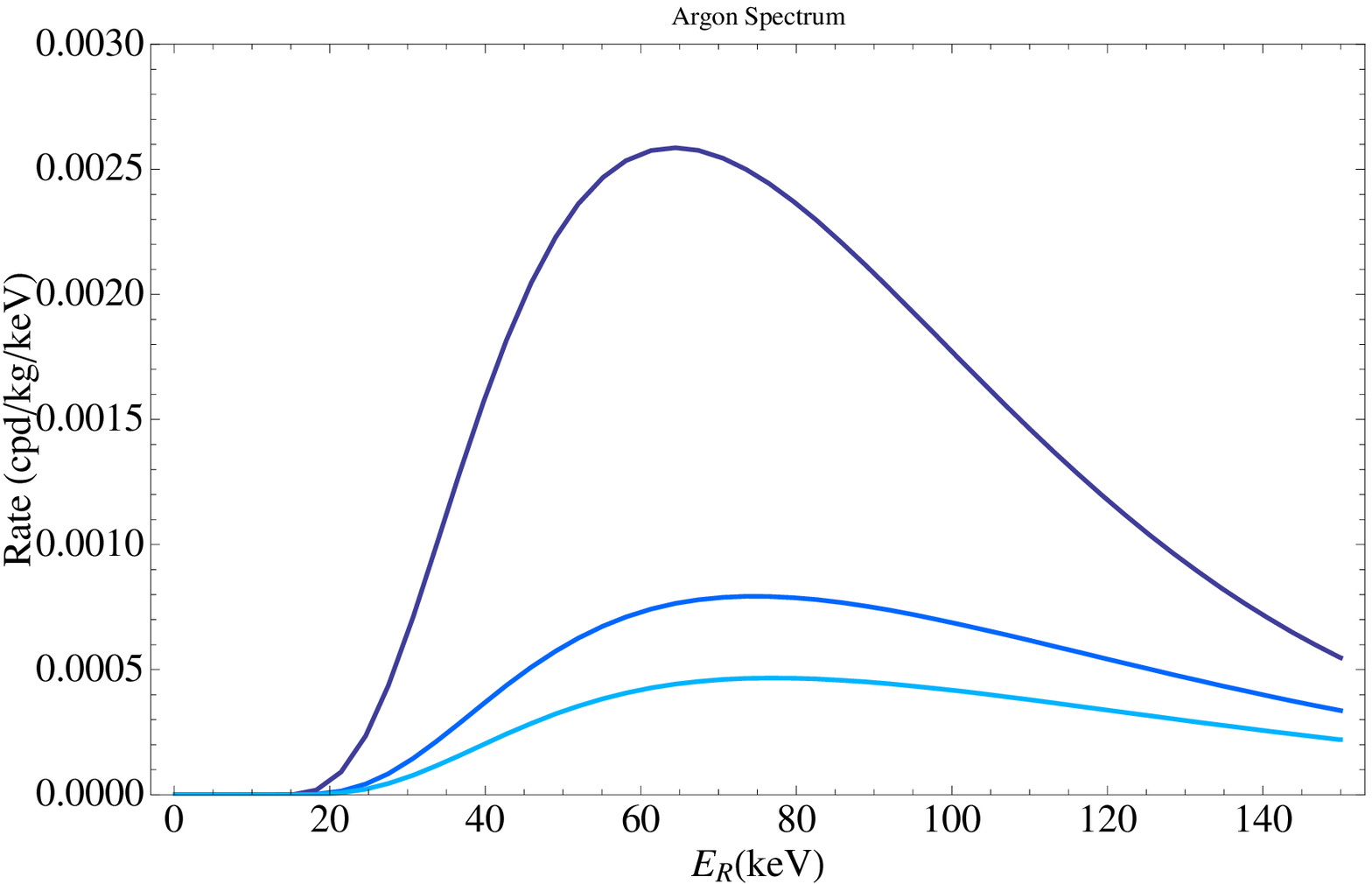}
\includegraphics[scale=.43]{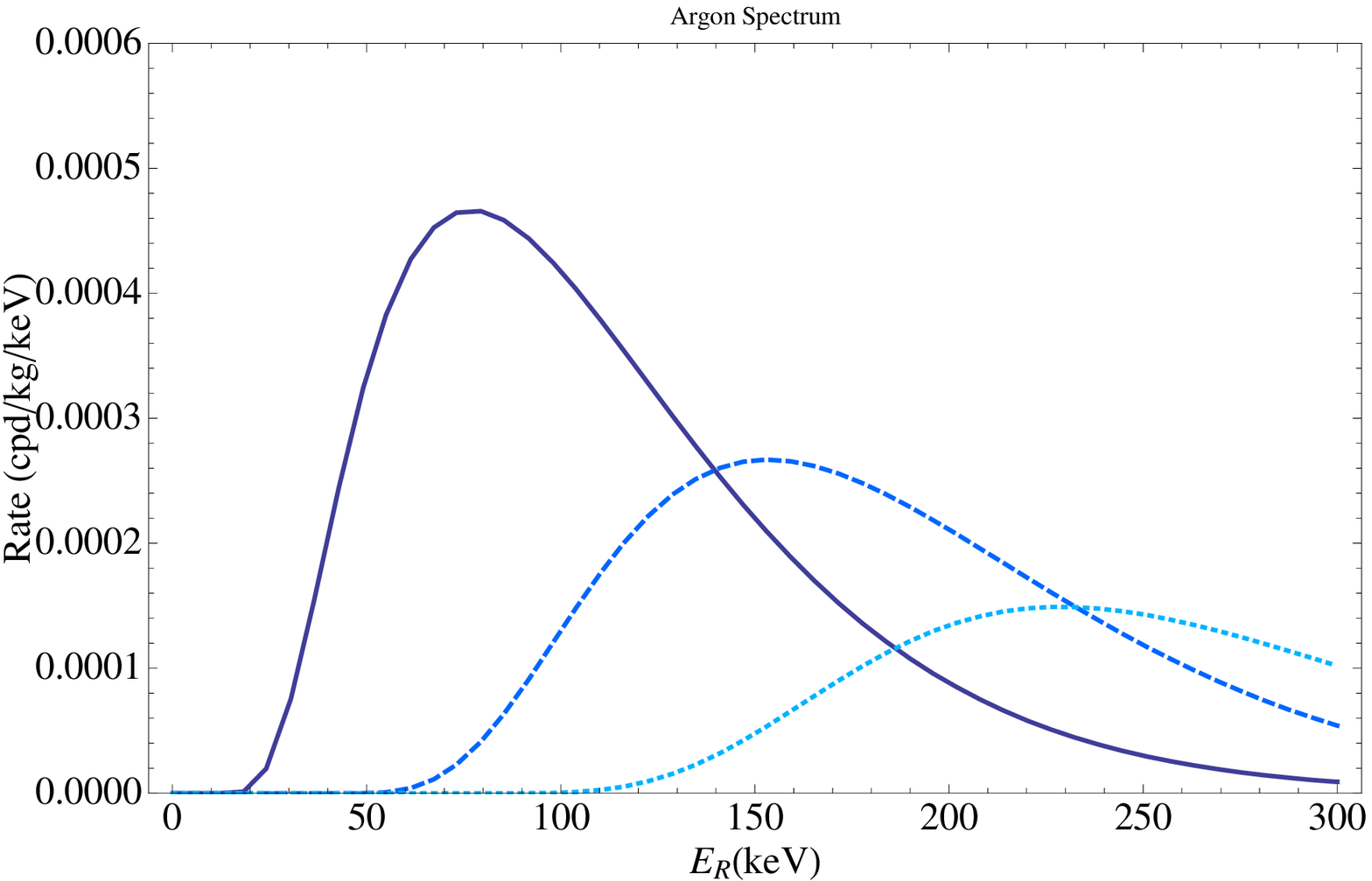}
\end{center}
\caption{In the left pane we show the Argon spectrum for $\delta = - 100$ keV  with WIMP-nucleon cross-section $\sigma = 10^{-42}~\mathrm{cm}^2$. The WIMP mass is 100 GeV, 300 GeV, and 500 GeV, from top to bottom. On the right, we fix the WIMP mass at 500 GeV and plot the resulting spectrum for $|\delta|=100$ GeV (solid), $|\delta|=200~{\rm GeV}$ (dashed), and $|\delta|=300$ GeV (dotted). As $|\delta|$ increases, the spectrum drifts outside the typical recoil energy observational window.}
\label{fig:ArSpec}
\end{figure}

\begin{figure}
\begin{center}
\includegraphics[scale=.53]{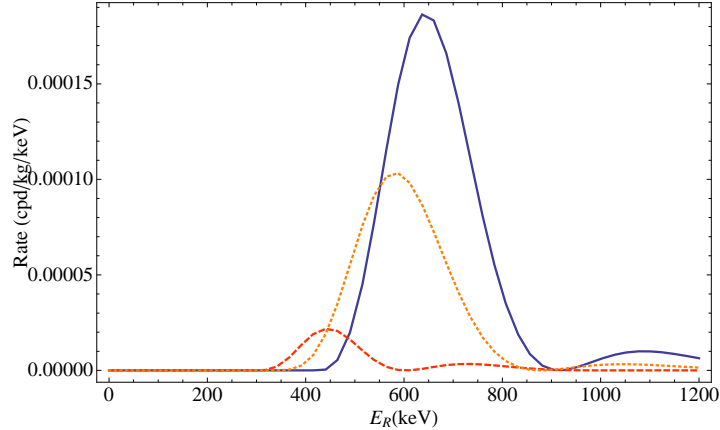}
\end{center}
\caption{The spectrum for $\delta = - 900$ keV and a WIMP mass of $500~{\rm GeV}$ with WIMP-nucleon cross-section $\sigma = 10^{-40}~\mathrm{cm}^2$. The targets are argon (solid-blue), germanium (dotted-orange), and xenon (dashed-red). Notice that in the case of deexcitation, targets with lighter elements actually yield larger rates.}
\label{fig:deex900keV}
\end{figure}

\begin{figure}
\begin{center}
\includegraphics[scale=.53]{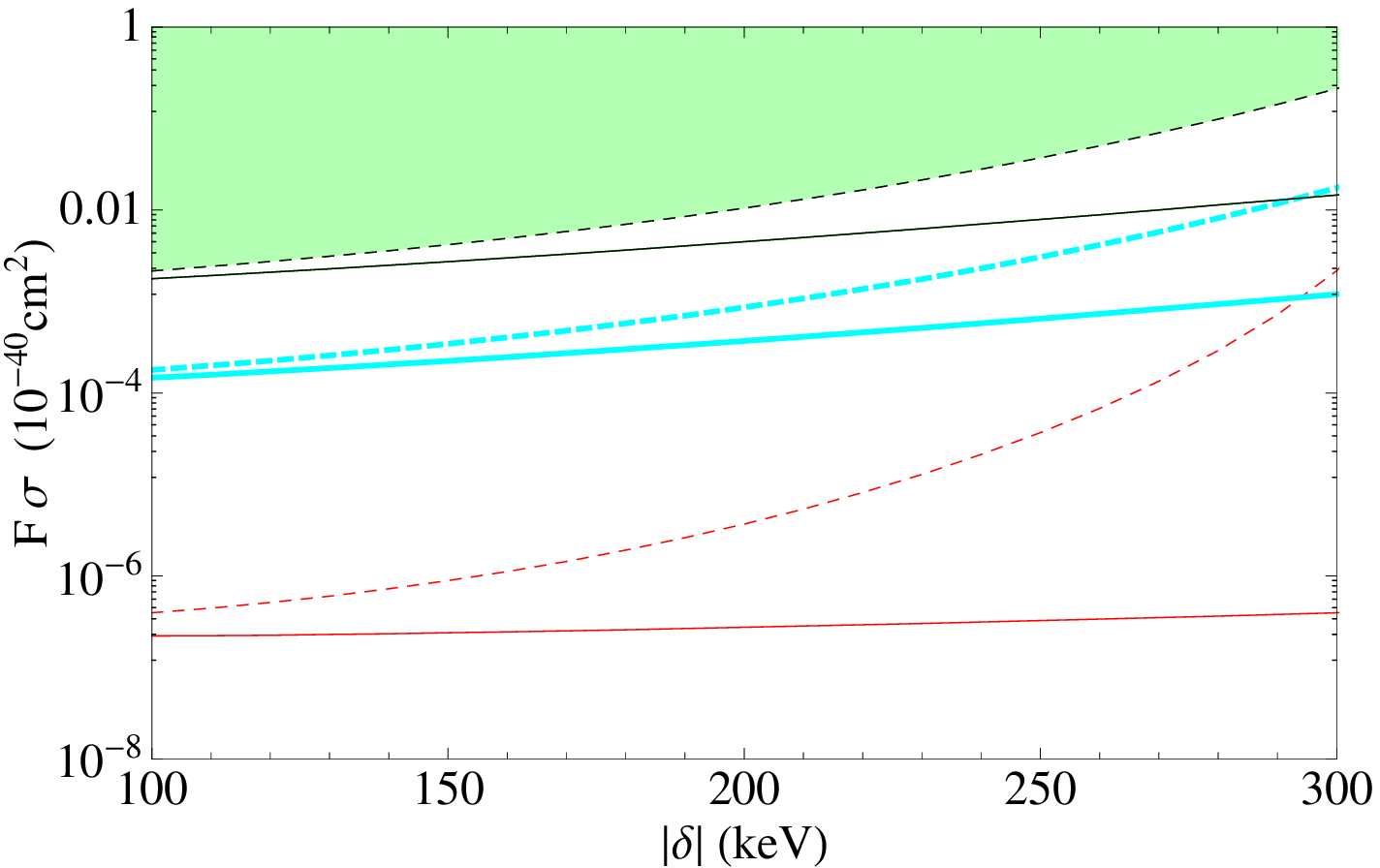}
\includegraphics[scale=.53]{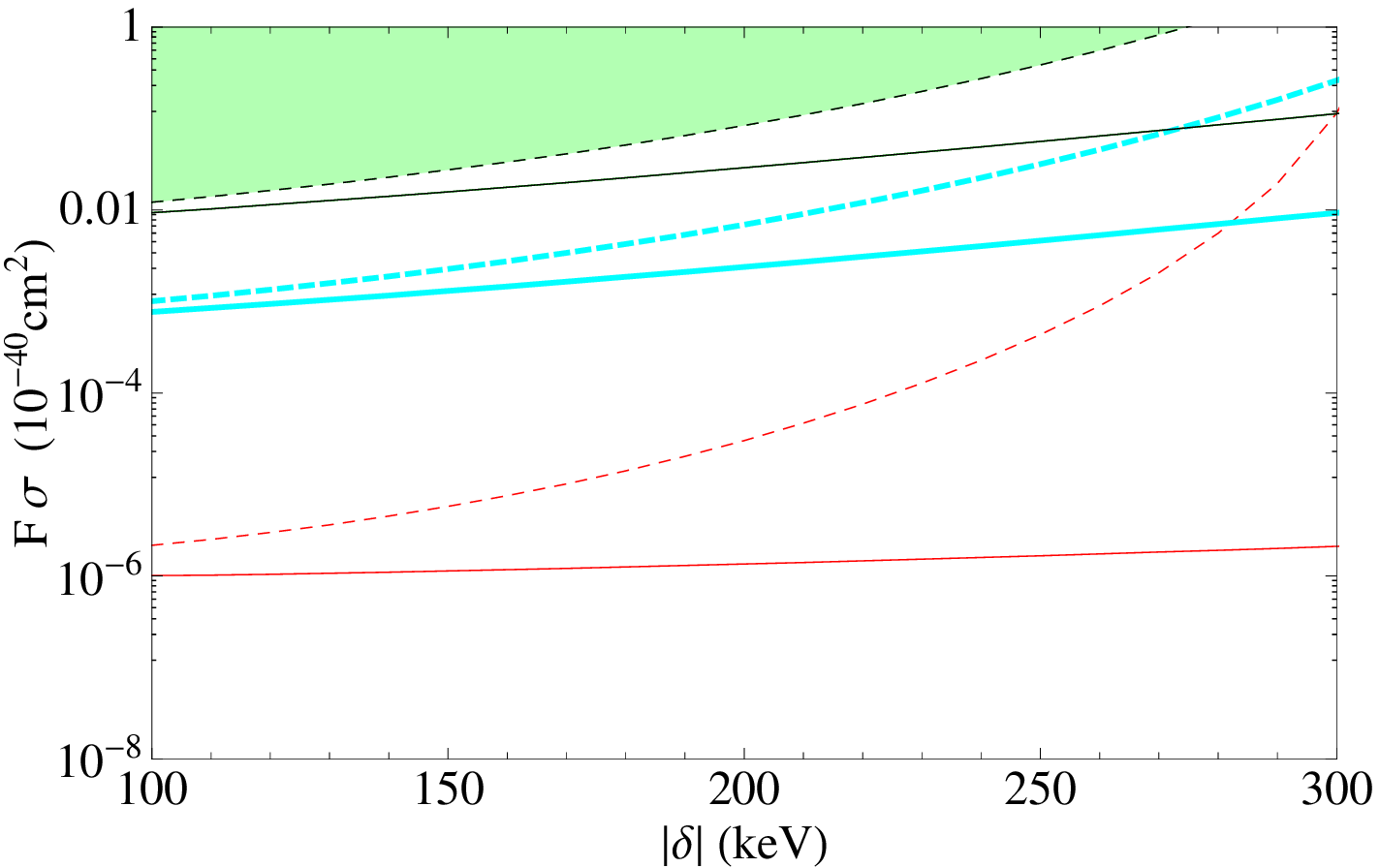}
\end{center}
\caption{The cross-section per nucleon $\times$ the fraction of excited state in the halo, $F$, is plotted against the splitting, $|\delta|$. The WIMP mass is fixed at 200 (1000) GeV  on the left (right) pane.  The shaded area is excluded by CDMS and hence a fraction $F \sim 10^{-2}-10^{-3}$ is required if we assume a cross-section of $\sigma = 10^{-40}{\rm cm}^2$ as required by DAMA. Discovery curves for the planned CDMS $4^{th}$ run and Argon target experiments are shown for exposures of $400$ kg-day and $1000$ kg-year respectively \cite{Arisaka200963}. The dashed curves assume a recoil energy window  between $10-100$ keV for CDMS and between $35-100$ keV for Argon, while the solid curves show the large increase in sensitivity if a window of $0-1000$ keV is utilized. The upper curves correspond to CDMS, the lower to Argon.}
\label{fig:deltaVssigma200}
\end{figure}

Experiments with Argon as the nuclear target promise to reach much larger exposures than currently available by either CDMS or XENON10. Since Argon is a lighter element than Germanium, the usual iDM scenario is generally invisible in such experiments. However, as discussed above, a metastable state can leave its mark by scattering off the Argon nucleus and deexciting into the ground state. In Fig. \ref{fig:ArSpec} we plot the recoil energy spectrum in Argon for $|\delta|=100$ keV and several choices of the WIMP mass. Also shown in that figure is the variation in the spectrum as $\delta$ is increased. The bulk of the signal is shifted to the right and might be missed altogether if the observational window is too restrictive. In order to contrast with other targets, in Fig. \ref{fig:deex900keV} we plot the spectrum resulting from a $900~{\rm keV}$ deexcitation against argon, germanium, and xenon. The lighter target experiments perform better in the case of deexcitation. In Fig. \ref{fig:deltaVssigma200} we present the discovery reach of argon and germanium in the $|\delta| - \sigma$ plane. At very large $|\delta|$, the experimental sensitivity is limited by the fact that most scattering events happen with a recoil energy which is outside the energy window (typically between $0 - 150$ keV). It is therefore important to broaden the search window to allow for nuclear recoils with higher energies. 

We note that at very large $|\delta|$, it is not clear that the form-factor really models the behavior of the scattering correctly, since the energy transfer is becoming comparable to nuclear excitation energies. However, the lowest lying excited state of $^{40}{\rm Ar}$ is at 1460 keV (excited states of  $^{36}{\rm Ar}$  and  $^{38}{\rm Ar}$ are higher). Since DM states with splittings $\delta \gsim 1$ MeV are cosmologically unstable, we do not expect excitations of this state and there should still be a clean signal of deexcitation in Argon experiments. This is also relevant for silicon targets, since the lowest lying excited state of $^{28}{\rm Si}$ is at 1780 keV (that of $^{29}{\rm Si}$ is only slightly lower).  Therefore, clean deexcitations may also be observed with Si targets and in fact we can expect 10's of such events with the current exposure of silicon at CDMS for a $500~{\rm GeV}$ WIMP with a $10^{-40}\cm^2$ scattering cross-section against nuclei. In contrast, future large exposures of Ge and Xe would be expected to yield significant rates of nuclear excitations, because of the many excited nuclear states which are accessible. Nevertheless, because these photons are monochromatic, they could be studied, rather than simply appearing as a background source.

Another possible signature a population of excited states can result in is the indirect detection of energetic neutrinos coming from the capture and subsequent annihilation of WIMPs in the sun or the earth. In the inelastic case of iDM, capture is more difficult because the WIMPs cannot scatter as efficiently anymore. In particular, considering that iron is lighter than germanium, the bounds from CDMS imply that it is not possible to capture WIMPs in the earth in the iDM scenario. This is no longer true if an abundance of excited states is present since they can scatter against any target. The resulting neutrino flux may be detected in upcoming neutrino telescopes experiments. 

Finally, we should note that these exothermic reactions can result in electron recoils \footnote{See \cite{Bernabei:2007gr,Bernabei:2008mv} for a relevant discussion involving light DM particles.}. These would fail cuts in standard WIMP searches, but could possibly show up in careful analyses \footnote{We thank Chris Stubbs for comments on this point.}. These electrons would have a spread of energies set by the range of binding energies of the atom in question, but would still be naturally narrow if the excited splitting were sufficiently high.

\section{Conclusions}
A wide range of experimental results compel us to consider dark matter with $\sim$ MeV excited states. Explaining PAMELA with new gauge forces requires new states of dark matter, and the effectiveness of the Sommerfeld enhancement requires they have mass splittings not much greater than the kinetic energy of a typical WIMP in the halo. Beyond this, we have other significant motivations for excited states from completely different signals. In particular, the presence of a $\sim 100~{\rm keV}$ excited WIMP state can reconcile DAMA with other experiments through inelastic collisions with a nucleus, while a $\sim 1~{\rm MeV}$ excited state can produce the INTEGRAL signal from WIMP-WIMP collisions in the galactic center. While states with $\delta \gsim 2 m_e$ are expected to be cosmologically unstable, those with smaller splittings can naturally have lifetimes longer than the age of the universe. However, it is important to keep in mind that simple alterations to the models can lead to more rapid decays.

If the excited states are indeed long-lived then the interactions in the dark sector can yield a relic fraction of excited states spanning many orders of magnitude, from $\mathcal{O}(1)$ to negligibly small. If such states exist, the downscattering signal for Ge, Xe, Si and Ar would be remarkable. $\delta \sim 100~{\rm keV}$ states would be visible at all experiments, while $\delta \sim 1~ {\rm MeV}$ would be easily visible at argon and silicon experiments. For Ge and Xe, the scatterings would produce large numbers of nuclear excitations, with associated photons. Coherent nuclear scatterings would lie outside of presently considered energy ranges, motivating a study of higher ($E_R \gsim 100~{\rm keV}$) energies.

Models with metastable excited states can change our interpretation of the INTEGRAL signal. For instance, if the DAMA transition is associated with an upscattering from a metastable state $\sim 900~{\rm keV}$ above the ground state to a state $\sim 1~{\rm MeV}$ above the ground state, then the INTEGRAL signal can be generated by WIMPs much lighter than 500  GeV. Such models naturally have an associated down-scattering signal as described above. Similarly, it is possible that the INTEGRAL signal is generated from collisions of metastable WIMPs where one is up-scattered and one is down-scattered. Such models also expand the parameter space for lighter mass XDM models, but do not obviously admit an interpretation of the DAMA data.

Such scenarios show the continued richness of possible signals at upcoming direct detection experiments. It is important to control backgrounds over the largest energy range possible, and to consider signals which peak at high energies, in addition to conventional WIMP scatters which have an exponentially falling spectrum. The next few years will significantly probe the parameter space of excited WIMP states associated with these higher energy scattering events.

\vskip 0.2in
{\bf \noindent Note added:} While this work was ongoing, \cite{Chen:2009dm} appeared, which also considered the inverted spectrum for XDM discussed in section \ref{subsec:inverted}, although without connection to iDM or nuclear scattering.

\vskip 0.2in
\noindent {\bf Acknowledgements}
The authors thanks Blas Cabrera, Clifford Cheung, Rick Gaitskell, Cristiano Galbiati, Lisa Goodenough, David E. Kaplan, Dan McKinsey,  Philip Schuster, Chris Stubbs, and Natalia Toro for useful discussions. IY thanks Maxim Pospelov for early and provocative conversations on the lifetimes of the excited states. The authors are especially indebted to Nima Arkani-Hamed, in collaboration with whom the analysis on lifetimes and freezeout was done, and who further provided many additional insights.
NW is supported by NSF CAREER grant PHY-0449818 and DOE OJI grant \# DE-FG02-06ER41417.  IY is supported by the NSF under grant PHY-0756966 and the DOE under grant DE-FG02-90ER40542.

\bibliographystyle{JHEP}
\bibliography{metastable}
\end{document}